\documentclass[10pt,a4paper,twoside]{article}
\usepackage{epsfig}
\usepackage{baltlat6}
\usepackage{array}
\usepackage{here}
\begin{document}
\ \
\vspace{0.5mm}
\setcounter{page}{277}

\titlehead{Baltic Astronomy, vol.\,XX, XXX--XXX, 2012}

\titleb{THE EFFECT OF AN INERT SOLID RESERVOIR ON MOLECULAR ABUNDANCES IN DENSE INTERSTELLAR CLOUDS}

\begin{authorl}
\authorb{Juris Kalv$\mathrm{\bar{a}}$ns}{1} and
\authorb{Ivar Shmeld}{1}
\end{authorl}

\begin{addressl}
\addressb{1} {Ventspils International Radio Astronomy Center of Ventspils University College, In\v{z}enieru iela 101, Ventspils, Latvia, LV-3601; kalvans@lu.lv, ivarss@venta.lv}

\end{addressl}

\submitb{Received: 2012 February 16; accepted: 2012 November 11}

\begin{summary} The question, what is the role of freeze-out of chemical species in determining the molecular abundances in the interstellar gas is a matter of debate. We investigate a theoretical case of a dense interstellar molecular cloud core by time-dependent modelling of chemical kinetics, where grain surface reactions deliberately are not included. That means, the gas-phase and solid-phase abundances are influenced by gas reactions, accretion on grains, and desorption, only. We compare the results to a reference model, where no accretion occurs and only gas-phase reactions are included.
We can trace that the purely physical processes of molecule accretion and desorption have major chemical consequences on the gas-phase chemistry. The main effect of introduction of the gas-grain interaction is long-term molecule abundance changes that come nowhere near an equilibrium in during the typical lifetime of a prestellar core.
\end{summary}


%% \resthead is the RUNNING TITLE at top of the pages
\resthead{The effect of inert molecule reservoir on interstellar abundances} {J. Kalv{\-a}ns, I. Shmeld}

\sectionb{1}{INTRODUCTION}

The observed abundances of various species in the interstellar medium are frequently explained by gas-phase reactions. However, in dense and dark clouds elements other than H and He mostly accrete on dust-grains, forming icy mantles. There they may undergo chemical reactions and photoprocessing, while still in an adsorption-desorption equilibrium with the gas phase. The extent and significance of the surface reactions are matter of debate (Woodall et al. 2006). Models that include grain surface-phase chemistry in addition to gas-phase chemistry e.g. Hasegawa et al. (1992), Allen \& Robinson (1977), generally have better and wider possibilities to reproduce the production of various species. However, the solid phase itself acts as a reservoir, which withdraws molecules from active gas phase chemical transformations. We aim to investigate the effect of the physical freeze-out of molecules onto grains on gas-phase chemical processes. A theoretical case is examined by chemical kinetics modelling, where the freeze-out of heavy elements occurs but the surface reactions are not included, with $\mathrm{H_2}$ production on grains as the only exception. Therefore, in addition  to the  gas-phase  reactions  the  accretion and desorption (Sect. 2.1) of species onto and from the grains are taken into account, only.

\sectionb{2}{THE MODEL}
To investigate the effect of an inert reservior on  molecular abundances we compare the results of two chemical kinetic models: Model 1 with gas phase species taken into account only, and  Model 2, where some of the neutral species are assumed refractory and can stick to grain surfaces, and the frozen grain mantle forms. The  model is essentially a time-dependent and simplified version of the  model described by  Kalvans \& Shmeld (2010) and Kalvans \& Shmeld (2012) -- a plane-parralel interstellar cloud, illuminated from both sides by the interstellar ultraviolet and cosmic-ray radiation field. However, we examine only a deep cloud core, where the visual absorption is 20 magnitudes, i.e. the interstellar ultraviolet radiation plays essentially no role. We use the UDFA06 dipole (Woodall et al. 2006) kinetic chemical reaction set. The gas temperature is taken 15K, dust temperature 10K. We assume atomic initial abundances. The calculations are performed with $3.0times10^{11}$ s as the initial time in 15 time steps with 1.4 as the multiplying factor until $3.3times10^{13}$ s is reached as the final time. For Model 2 accretion on dust grains and desorption of molecules but no surface or other solid-phase reactions are included. Of the 301 chemical species included 83 are assumed refractory and form a mantle. The abundance and depletion of the elements for the model is shown in Table 1. We opted for a case, when elements are already heavily depleted, since the case of an already formed molecular cloud core is investigated.

\begin{table}[!t]
\begin{center}
\vbox{\small\tabcolsep=6pt
\parbox[c]{124mm}{\baselineskip=10pt
{\normbf\ \ Table 1.}{\norm\
The adopted abundances and depletion of elements used in the model.
\lstrut}}
\begin{tabular}{c|cccc}
\hline
 &Total& & &Initial\\
Element&abundance$^{\mathrm{a}}$&Depletion,&Source&gas-phase\\
 & & \% & &abundance$^{\mathrm{a}}$\\
\hline
H&1.0&0&--&1.0\\
He&0.1&0&--&0.1\\
C&2.9E-04&39&J09$^{\mathrm{c}}$&1.8E-04\\
N&7.9E-05&22&J09&6.2E-05\\
O&5.8E-04&42&J09&3.3E-04\\
Mg&4.2E-05&95&J09&2.3E-06\\
Fe&3.5E-05&99&J09&2.0E-07\\
S&2.1E-05&92&J09, P97$^{\mathrm{d}}$&1.8E-06\\
Na&2.1E-06&75&P97, PPG84$^{\mathrm{e}}$&5.3E-07\\\hline
\end{tabular}
}
\end{center}
\begin{list}{}{}
\item[$^{\mathrm{a}}$] relative to hydrogen
\item[$^{\mathrm{b}}$] Prodanovic et al. (2010)
\item[$^{\mathrm{c}}$] Jenkins (2009), with $F_*=1$
\item[$^{\mathrm{d}}$] Pagel (1997)
\item[$^{\mathrm{e}}$] Phillips et al. (1984)
\end{list}
\vskip-4mm
\end{table}

\subsectionb{2.1}{Accretion and desorption}
The rate coefficients for accretion of molecules onto grains (Willacy \& Williams 1993), thermal evaporation and desorption by grain heating by cosmic rays (Hasegawa \& Herbst 1993), desorption by cosmic-ray induced photons (Prasad \& Tarafdar 1983, Willacy, Williams 1993), desorption of surface molecules by energy released by $\mathrm{H_{2}}$ molecule formation on grains (Roberts et al. 2007), and direct ejection of molecules from the mantle by cosmic-rays all are calculated according to Kalvans \& Shmeld (2010). The species desorbed by $\mathrm{H_{2}}$ formation are taken those with binding energies $E_{b} \leq 1210K$ (binding energies from Aikawa et al. 1997).

\subsectionb{2.2}{The formation of the mantle}
The mantle-surface model is rather simple and was made to include the two phases in a way that in the period modelled they essentially accrete with only minor return to the gas phase. The mathematical model is such that the surface molecules undergo various desorption processes (Sect. 2.1) and turn into mantle molecules with a constant rate, while the mantle molecules may undergo direct ejection into gas phase and return to surface. First, the rate coefficient of mantle molecules is calculated by

   \begin{equation}
   \label{dig-up}
k_{dig-up}=t^{-1}_{cr}\times v^{-1},
   \end{equation}

where $t_{cr}=10^{12}$ (Bringa \& Johnson 2004) is the average time between two successive strikes by Fe cosmic-ray nuclei and \textit{v} is the assumed number of strikes needed to return all mantle molecules to the outer surface for some time. Then, the rate coefficient of surface molecules turning into mantle molecules is assumed to be $k_{mantle}=100k_{dig-up}$ and it would produce a mantle-to-surface total abundance ratio of 100:1.

We choose \textit{v}=5000, so that the time to reach the approximate 100:1 ratio is on the order of 1 Myr, more or less consistent with the time of mantle compression, as found by Palumbo (2006) and Accolla et al. (2011). A high abundance of molecules exposed to outer surface (low mantle-to-surface ratio) means a high porosity of the mantle. The porosity is reduced over time by cosmic-ray hits, exothermic reactions and other events (see Palumbo, 2006, Accolla et al., 2011, Raut et al., 2007a, b, 2008).

This mathematically simple picture is the same as used in Kalvans \& Shmeld (2012). It is not quite physically adequate, and is a little more sophisticated than the one used in Kalvans \& Shmeld (2010). However, it includes the accretion process and interaction of the solid grain mantle with gas phase and allows us to study the effect of the introduction of an inert reservior on the gas-phase abundances of chemical species.

%The selected coefficients ensure that approximately 99\% of all molecules are in solid phase after a time of $3\times10^{13}$ s.

\sectionb{3.} {RESULTS AND DISCUSSION}

The results are graphically presented in Figs. 1 to 6. The calculated abundances for selected species for Model 1 are shown in Fig. 1, the gas, solid and total abundances for Model 2 are shown in Figs. 2, 3 and 4, respectively. For comparison, the ratio for Model 2 to Model 1 calculated gas and total abundances are shown in Figs. 5 and 6, respectively.

%%%%%%%%%%%%%%%%%%%%%%%%%%%%%%  FIGURE 1, M1, 2.5.

\begin{figure}[!tH]
\vspace{-8.5cm}
\vbox{
\centerline{\psfig{figure=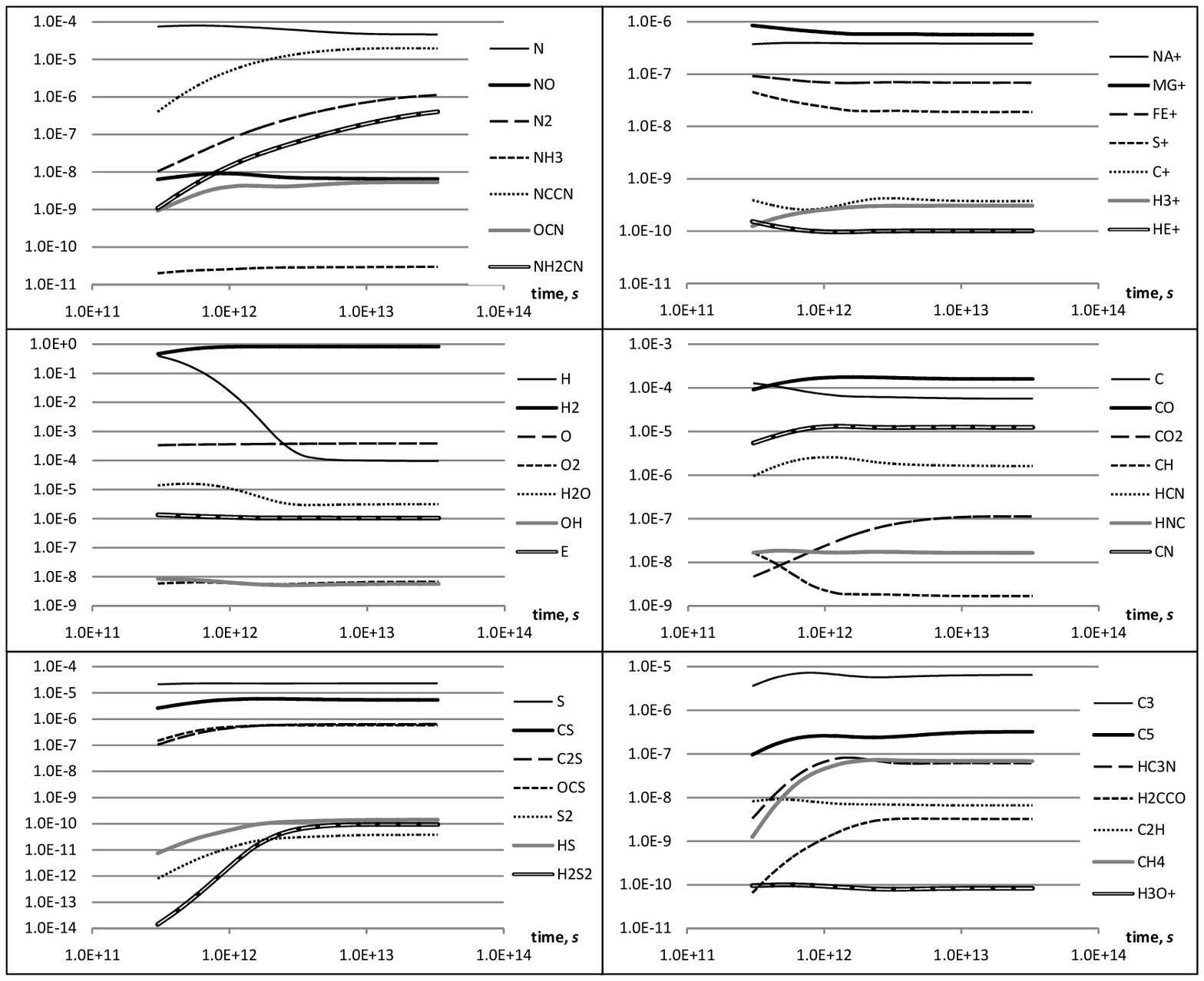, width=160mm,angle=0,clip=}}
\vspace{-1.5cm}
\captionb{1}
{Calculated gas-phase relative abundances for selected species for Model 1 (pure gas-phase). Pay attention to the relative abundance (ordinate) scale for each graph.}
}
\end{figure}

%%%%%%%%%%%%%%%%%%%%%%%%%%%%%%  FIGURE 2, M2 gas, 2.6.

\begin{figure}[!tH]
\vspace{-8.5cm}
\vbox{
\centerline{\psfig{figure=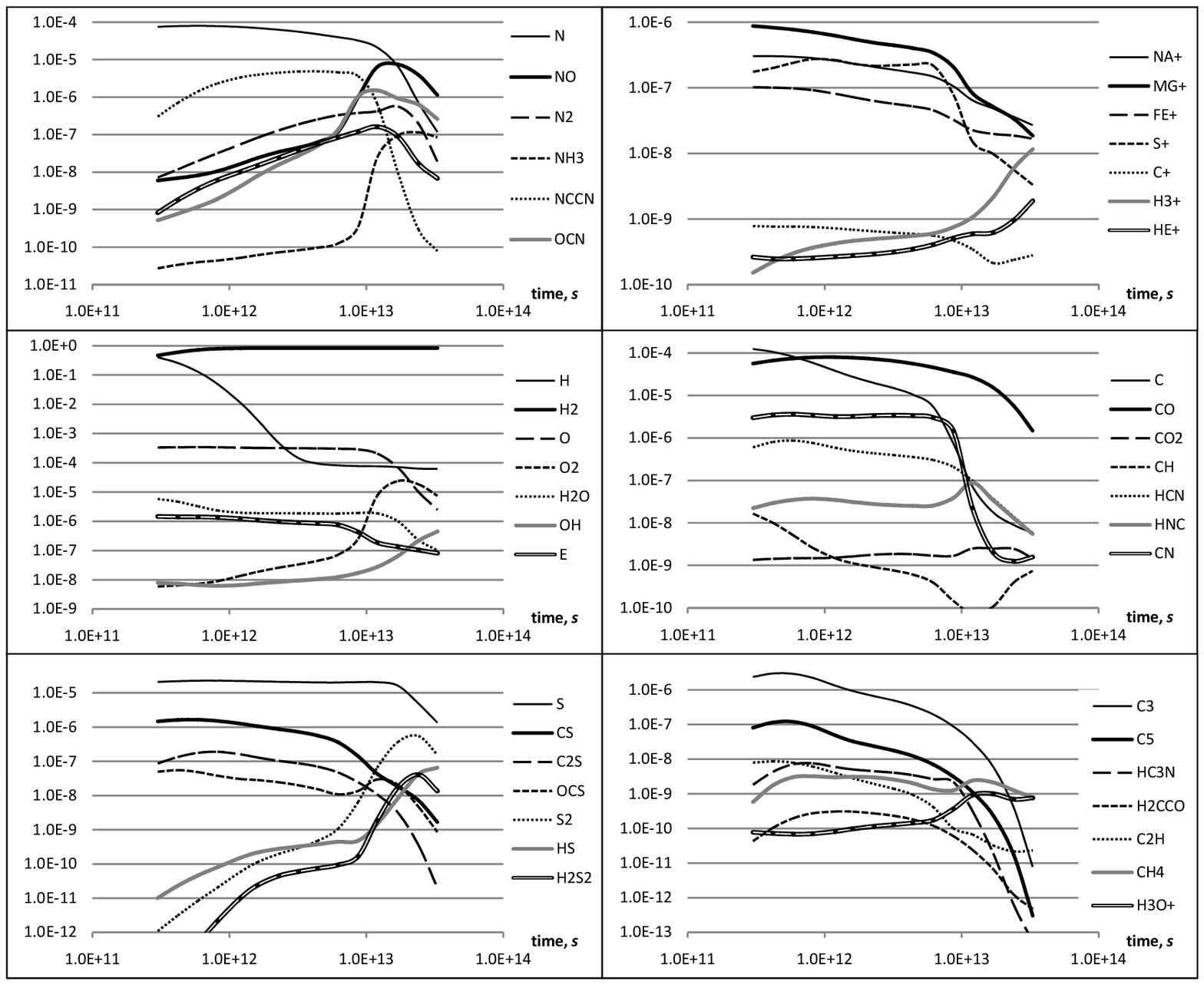, width=160mm,angle=0,clip=}}
\vspace{-1.5cm}
\captionb{2}
{Calculated gas-phase relative abundances for selected species for Model 2 (with freeze processes). Pay attention to the relative abundance (ordinate) scale for each graph.}
}
\end{figure}

%%%%%%%%%%%%%%%%%%%%%%%%%%%%%%  FIGURE 3, M2 solid, 2.8.

\begin{figure}[!tH]
\vspace{-12cm}
\vbox{
\centerline{\psfig{figure=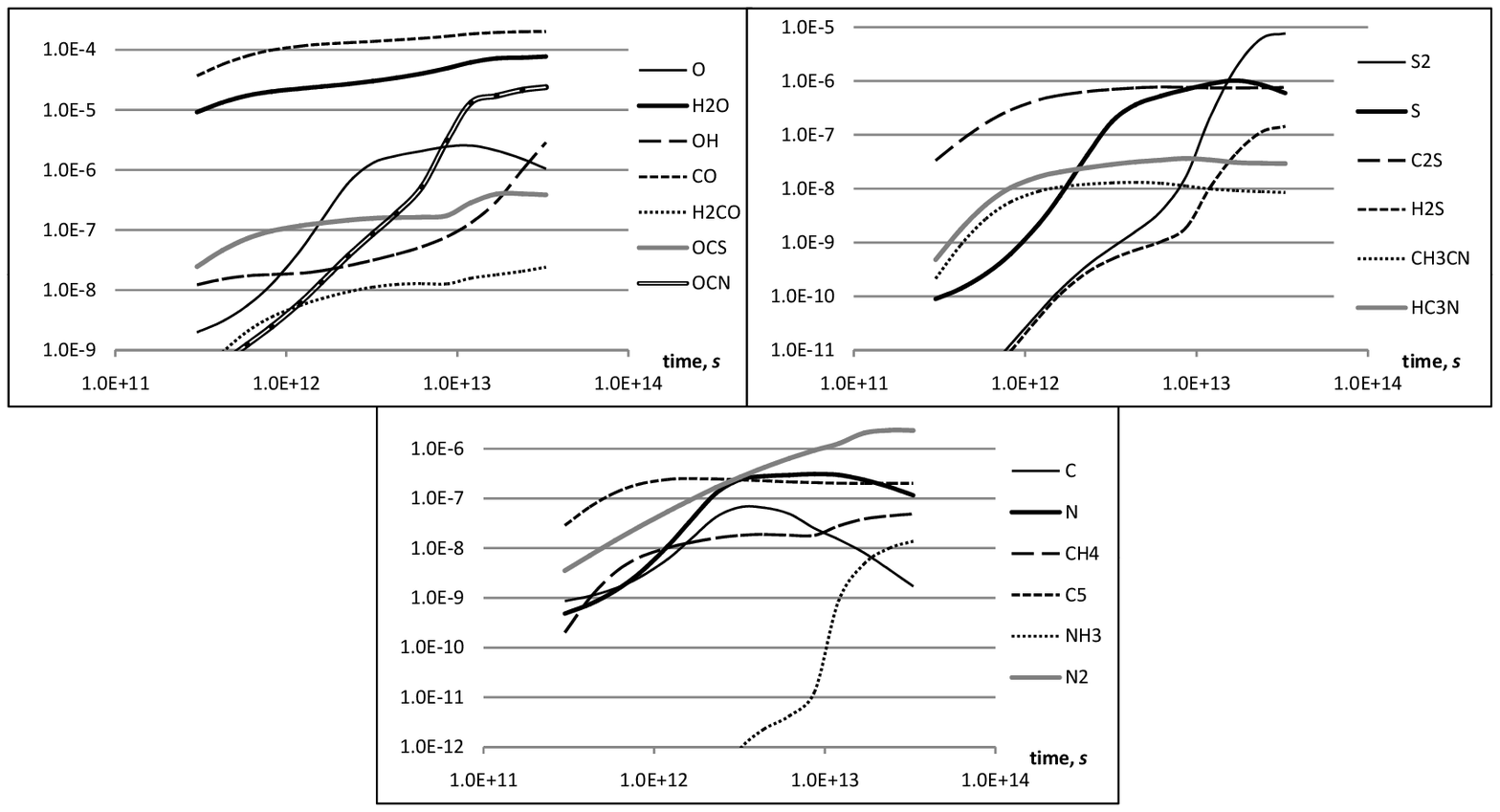, width=160mm,angle=0,clip=}}
\vspace{-1.5cm}
\captionb{3}
{Calculated solid-phase relative abundances for selected species for Model 2 (with freeze processes). Pay attention to the relative abundance (ordinate) scale for each graph.}
}
\end{figure}

%%%%%%%%%%%%%%%%%%%%%%%%%%%%%%  FIGURE 4, M2 total, 2.9.

\begin{figure}[!tH]
\vspace{-12cm}
\vbox{
\centerline{\psfig{figure=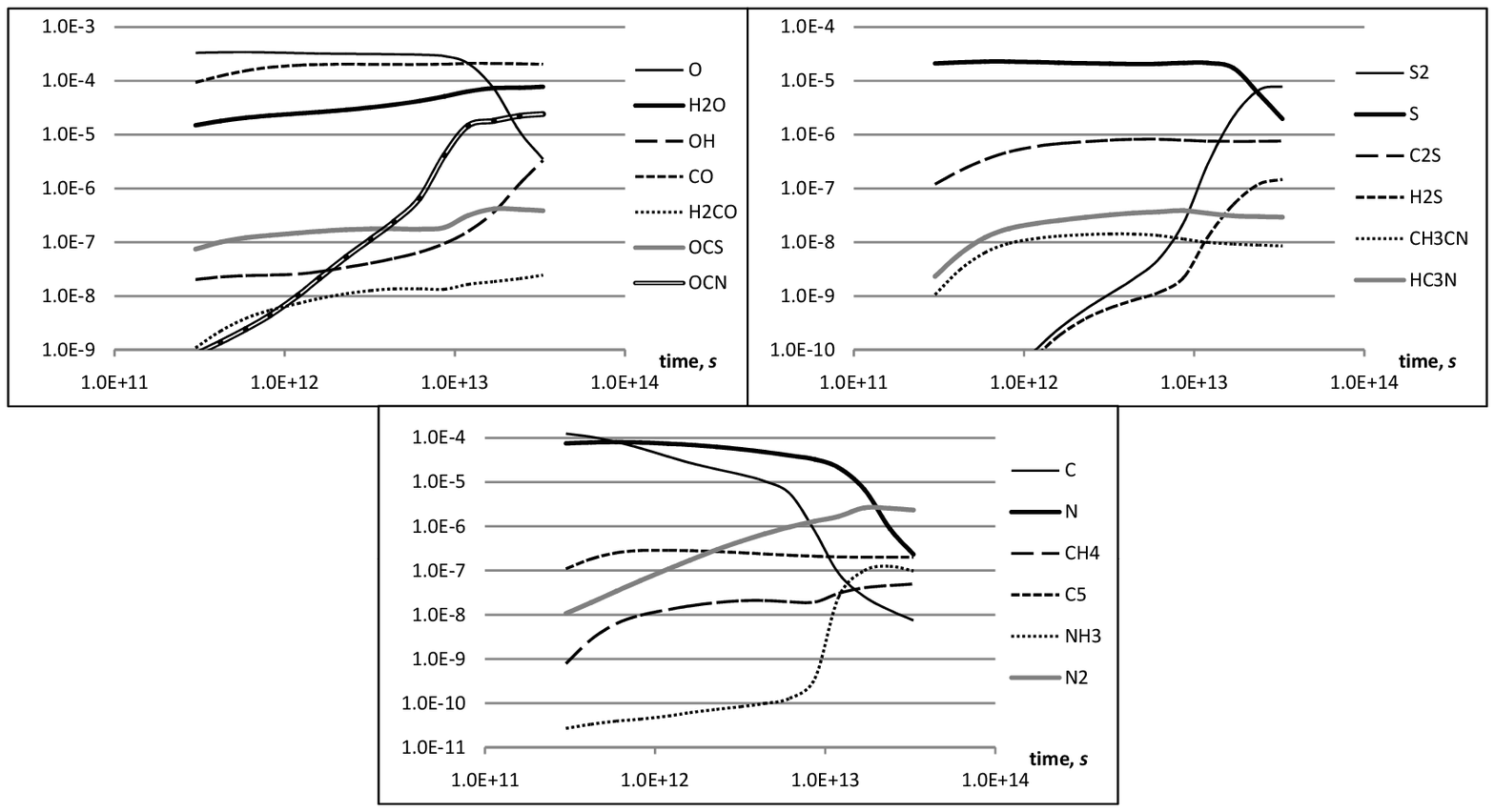, width=160mm,angle=0,clip=}}
\vspace{-1.5cm}
\captionb{4}
{Calculated total (solid + gas phase) relative abundances for selected species for Model 2 (with freeze processes). Pay attention to the relative abundance (ordinate) scale for each graph.}
}
\end{figure}

%%%%%%%%%%%%%%%%%%%%%%%%%%%%%%  FIGURE 5, M2/M1 gas, 2.10.

\begin{figure}[!tH]
\vspace{-12cm}
\vbox{
\centerline{\psfig{figure=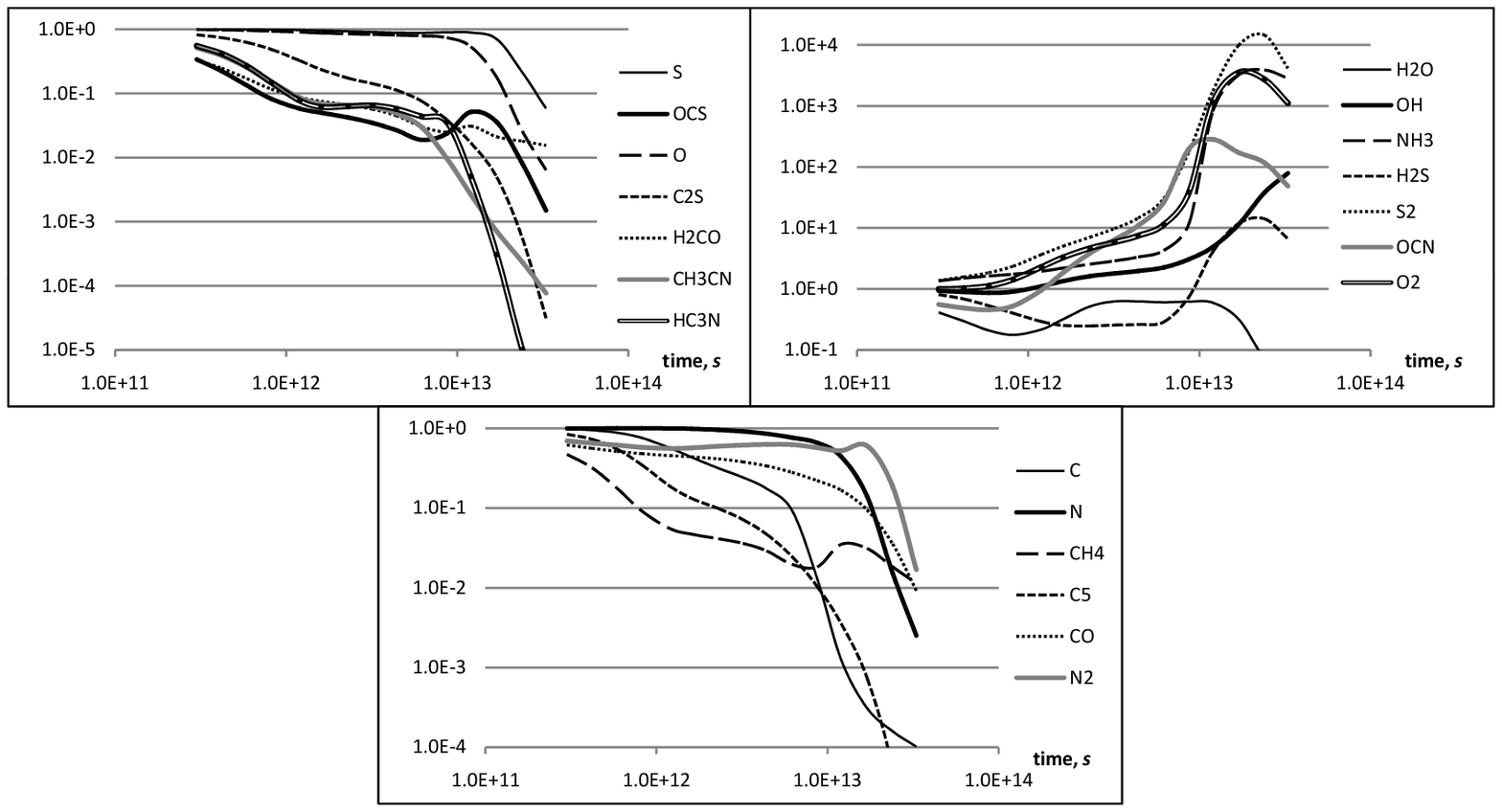, width=160mm,angle=0,clip=}}
\vspace{-1.5cm}
\captionb{5}
{The comparison of calculated gas-phase abundances for selected species between Model 2 (with freeze processes) and Model 1. The ratio of Model 2 to Model 1 gas abundances is shown. Pay attention to the M2/M1 (ordinate) scale for each graph.}
}
\end{figure}

%%%%%%%%%%%%%%%%%%%%%%%%%%%%%%  FIGURE 6, M2/M1 total, 2.11.

\begin{figure}[!tH]
\vspace{-12cm}
\vbox{
\centerline{\psfig{figure=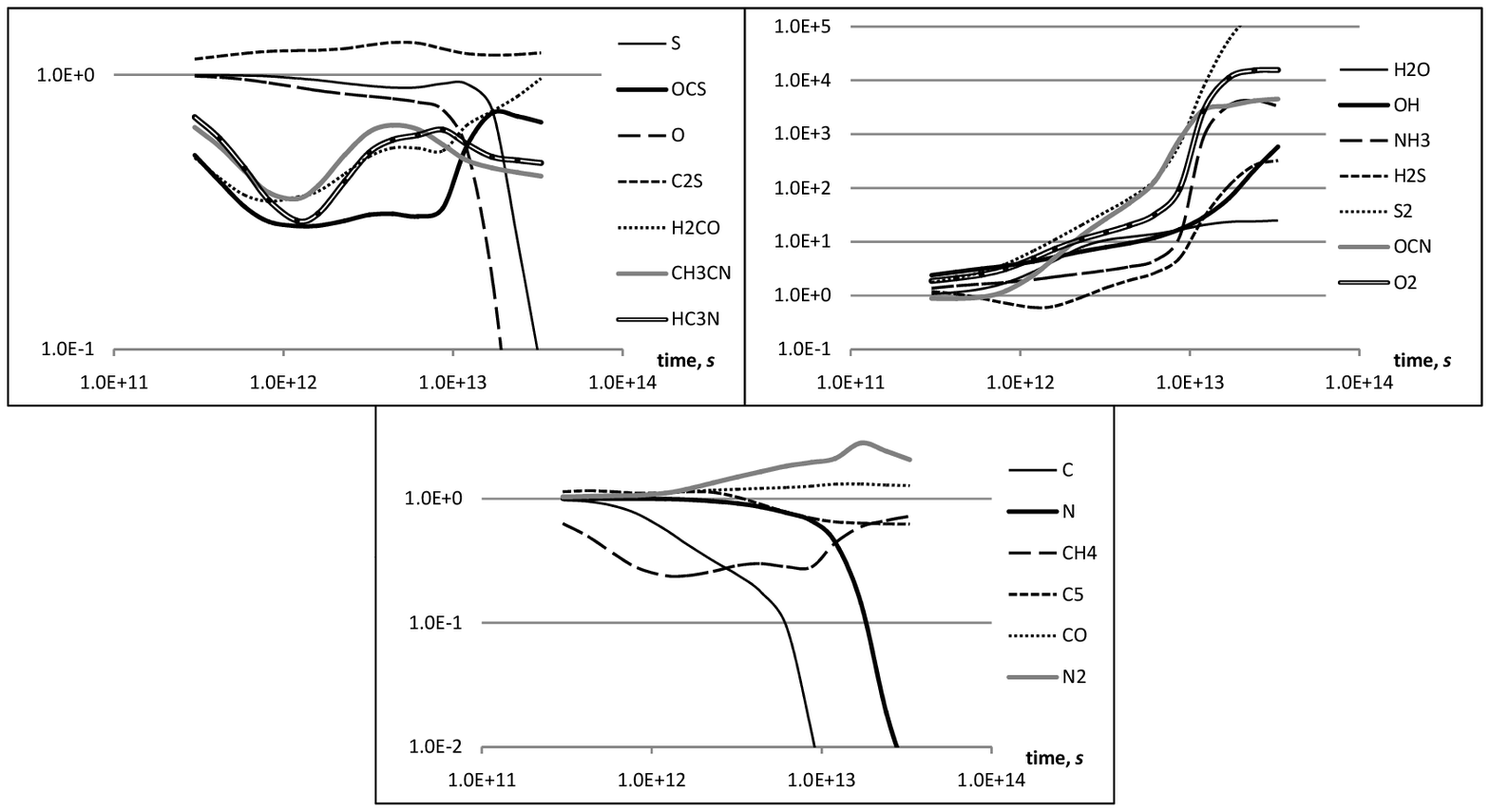, width=160mm, angle=0,clip=}}
\vspace{-1.5cm}
\captionb{6}
{The comparison of calculated total (gas + solid phase) abundances for selected species between Model 2 (with freeze processes) and Model 1. The ratio of Model 2 to Model 1 total abundances is shown. Pay attention to the M2/M1 (ordinate) scale for each graph.}
}
\end{figure}

%%%%%%%%%%%%%%%%%%%%%%%%%%%%%%

The model with constant physical parameters approximately represents a case of a slow-contracting molecular core. Chemical reactions compete with the accretion of molecules on grains. Small molecules form and freeze faster than more complicated and heavy species. However, they have smaller adsorbtion energies and more easily desorb to the gas-phase. At time 1 Myr $>$ 95 \% of the heavy elements are in solid phase and the freeze-out continues, consistent with models by Brown \& Charnley (1990), Hasegawa et al. (1992) and Roberts et al. (2007).

The molecule accretion process facilitates the transformation of free atoms into molecules. In Model 1 atoms like C, O, N and S are the major species along with the CO molecule (Fig. 1), while in Model 2 molecules like CO, O$_2$, H$_2$O, CN, NO, S$_2$ etc. take over, especially so, when one takes into account the total (gas + solid phase) abundances (Figs. 2, 3 and 4). Simple molecules with usually 2 metal atoms contain the majority of heavy elements. The largest \textit{increase} of abundances for Model 2 compared to Model 1 at 1 Myr is for molecules with 1-3 metal atoms with varying degrees of hydrogen content. The abundance increase for multi-atom carbon-chain species is usually only moderate and they have become depleted from the gas-phase. These results make it difficult to explain high abundance of these molecules in quiescent clouds without grain reactions (as in Herbst \& Leung 1989). As the accretion-desorption equilibrium approaches, the total abundance of many important small molecules (CO, CN, N$_2$, CS, etc.) begins to slowly become smaller. The atoms are then distributed to species with longer formation and freeze timescales, including multi-atom species. Thus, the chemical diversity for long lived clouds can be enhanced with time by this process.

Except for a few molecules (N$_2$, NH$_2$CN) the change of concentration for Model 1 molecules has nearly ceased already at 0.5 Myr, indicating a state close to a chemical equilibrium. It is very unlike Model 2, where both, gas and solid phase abundances are still considerably changing at 1 Myr. One can conclude that the physical gas-grain interaction can cause lasting chemical effects that may allow to deduce the history of an observed molecular cloud core or young stellar object. However, for a credible investigation a full interstellar gas-grain physics and chemistry model is needed.

\thanks
{The authors are thankful to Institute of Astronomy, University of Latvia for providing technical support for the calculations performed for this paper.}

\References

\refb  Accolla, M., Congiu, E., Dulieu, F. et al. 2011, Phys. Chem. Chem. Phys., 13, 8037

\refb  Aikawa, Y., Umebayashi, T., Nakano, T., Miyama, S. M. 1997, ApJ, 486, L51

\refb  Allen, M., Robinson, G. W. 1977, ApJ, 212, 396

\refb  Bringa, E. M. \& Johnson, R. E. 2004, ApJ, 603, 159

\refb  Brown, P. D. \& Charnley, S. B. 1990, MNRAS, 244, 432

\refb  Charnley, S. B., Rodgers, S. D., Ehrenfreund P. 2001, A\&A, 378, 1024

\refb  Goldsmith, P. F., Liseau, R., Bell, T. A. et al. 2011, ApJ, 737, 96

\refb  Hasegawa, T. I., Herbst \& E., Leung, C. E. 1992, ApJS, 82, 167

\refb  Hasegawa, T. I., Herbst, E. 1993, MNRAS, 261, 83

\refb	 Herbst, E. \& Leung, C. M. 1989, ApJS, 69, 271

\refb  Jenkins, E. B. 2009, ApJ, 700, 1299

\refb  Kalvans, J., Shmeld, I. 2010, A\&A, 521, A37

\refb  Kalvans, J., Shmeld, I. 2012, A\&A, submitted

\refb  $\mathrm{\ddot{O}}$berg, K. I., Boogert, A. C. A., Pontoppidan, K. M. et al. 2011b, ApJ, 740, 109 

\refb  Pagel, B. E. J. 1997, Nucleosynthesis and Chemical Evolution of Galaxies, Cambridge University Press, 84

\refb  Palumbo, M. E. 2006, A\&A, 453, 903

\refb  Phillips, A. P., Pettini, M., Gondhalekar, P. M., 1984, MNRAS, 206, 337

\refb  Prasad S.S., Tarafdar S.P. 1983, ApJ, 267, 603

\refb  Raut, U., Fama, M., Loeffler, M. J., Baragiola, R. A. 2008, ApJ, 687, 1070

\refb  Raut, U., Fama, M., Teolis, B. D., Baragiola, R. A. 2007a, J. Chem. Phys., 27, 204713

\refb  Raut, U., Teolis, B. D., Loeffler, M. J., Vidal, R. A., Fama, M., Baragiola, R. A. 2007b, J. Chem. Phys., 126, 244511

\refb  Roberts, J. F., Rawlings, J. M. C., Viti, S., Williams,  D. A. 2007, MNRAS, 382, 733

\refb  Sakai, N., Yamamoto, S. 2011, in IAUS 280, The Molecular Universe, Proc. IAU Symp., 280, 43

\refb  Willacy, K., Williams, D. A. 1993, MNRAS, 260, 635

\refb  Woodall, J., Ag\'{u}ndez, M., Markwick-Kemper, A. J., Millar, T. J. 2006, A\&A, 466, 1197

\end{document}